\documentstyle[epsf]{article}
\title{
Numerical analysis of the magnetic-field-tuned
superconductor-insulator transition 
in two dimensions
}
\author{Yoshihiro Nishiyama \\
{\it Department of Physics, Faculty of Science,
Okayama University}\\
{\it Okayama 700-8530, Japan}}
\date{(Received \hspace*{50mm})}

\begin{document}

\maketitle



\section*{Abstract}
Ground state of the two-dimensional hard-core-boson model
subjected to external magnetic field and quenched random chemical
potential is studied numerically.
In experiments, magnetic-field-tuned superconductor-insulator
transition has already come under through investigation,
whereas
in computer simulation, 
only randomness-driven localization (with zero magnetic field)
has been studied so far:
The external magnetic field brings about a difficulty that
the hopping amplitude becomes complex number (through the gauge twist),
for which the quantum Monte-Carlo simulation fails.
Here, we employ the exact
diagonalization method, with which 
we demonstrate that the model does exhibit field-tuned localization transition
at a certain critical magnetic field.
At the critical point, 
we found that
the DC conductivity is not universal,
but is substantially larger than that
of the randomness-driven localization transition
at zero magnetic field.
Our result supports recent experiment
by Markovi\'c {\it et al.}
reporting an increase of the critical conductivity with 
magnetic field strengthened.

\vspace{1cm}

{\bf PACS}: 
 74.76.w Superconducting films,
 71.23.An Theories and models; localized states,
 75.40.Mg Numerical simulation studies,
 68.35.Rh Phase transitions and critical
          phenomena,
 75.10.Nr Spin-glass and other random models.

{\bf Keywords}: magnetic-field-tuned superconductor-insulator transition,
exact-diagonalization method, randomness,
critical conductivity, dynamical critical exponent,
finite-size-scaling method.


\section{Introduction}
\label{section1}

Scaling argument of Abrahams, Anderson, Licciardello and Ramakrishnan
\cite{Abrahams79}
states that in {\it two} dimensions, infinitesimal amount of quenched
randomness should drive itinerant extended states to localize.
That is, at absolute zero temperature, the conductivity should vanish,
if there exist any randomnesses.
There are, however, some exceptions where the above description
fails. 
Metal-insulator transition found in MOSFET devise 
\cite{Kravchenko94,Kravchenko95,Kravchenko96}, for instance, is
one of recent hot topics
arousing much attention.
In the above-mentioned scaling theory,
the randomness perturbation appears to be marginal so that some unexpected
factors, for example, many-body interaction and external magnetic field,
would possibly change the scenario.

Suppose that there exists an attractive interaction among particles.
At the ground state,the particles would be unstable against bose condensation
so that the system would fall into a superconducting state.
Localization transition from the superconducting state is apparently out of
the scope of the conventional localization theory, and has
been studied extensively so far:
In experiments, the transition is observed for metallic films 
\cite{Haviland89,Liu91,Lee90,Hebard85,Wang91}
and Josephson-junction arrays \cite{Geerligs89,Katsumoto95}.
In essence,
${}^4{\rm He}$ film adsorbed by porous media belongs to the same
physics as well
\cite{Crooker83,Finotello88,Crowell97}, although one cannot measure
its electronic conductivity.
One of the main concerns, particularly in experiments, would be
the conductivity at the localization point.
An argument claims that localization transition occurs at a universal
condition \cite{Fisher90b}. 
That is, at the transition point, irrespective of the samples
examined, the electrical conductivity stays
$\approx (2e)^2/h$ with the electron charge $e$ and the Plank constant $h$.
A great number of experiments are trying to validate this issue.
From theoretical view point,
the criticality itself is a matter of interest
\cite{Ma86,Fisher89,Fisher90a,Fisher90b}.
It is to be stressed that 
the localization transition occurs at the ground state (zero temperature).
Such ground-state criticality differs significantly
from that of the finite-temperature
transitions, and belongs to a peculiar universality class:
From the path-integral viewpoint for the partition function,
$d$-dimensional quantum system is regarded as a $(d+1)$-dimensional
classical system.
The system size along the imaginary-time direction is given by the inverse 
temperature, which diverges at zero temperature.
Critical fluctuation along the imaginary-time direction (quantum fluctuation)
contributes to the nature of critical phenomenon,
giving rise to
a new universality class.
This extra contribution is characterized by the dynamical critical exponent
$z$;
a formal scaling argument on the superconductor-insulator
transition will be found 
 in the literatures \cite{Fisher89,Fisher90a}.
We stress that such arguments on criticality
are not of mere theoretical interest, but are readily measurable in
experiments; see Table \ref{table1}.

In order to tune the localization transition,
either magnetic field strength or amount of randomness (film thickness)
has to be adjusted carefully.
Because the former is far more advantageous for fine tuning,
numerous experiments are devoted
to the magnetic-field-tuned 
localization transition.
In Table \ref{table1}, we presented a summary of experimental results.
Among them, we would like to draw reader's attention to the
recent experiment by
Markovi\'c {\it et al.} \cite{Markovic98,Markovic99}, 
who scanned both two tuning parameters fairly systematically.

While in experiments, such the subtle issues have already come
under through investigation, in computer simulation, on the contrary,
it has not yet been done to incorporate 
the external magnetic field:
So far, transitions driven solely by the random chemical potential
(with zero magnetic field) have been simulated 
\cite{Runge92,Makivic93,Batrouni93,Wallin94,Sorensen92,Zhang95,Nishiyama99}.
Implementation of the magnetic field brings about a complication
such that the hopping amplitude becomes complex number.
Because of this,
quantum Monte-Carlo simulation does not work;
the inefficiency would be even more serious than 
that of the so-called 
negative sign problem.
(Note that here, the magnetic field is {\em not} meant to couple to 
the moment of spin
(as in usual spin models), but it rather
couples to the charge (hopping) through the gauge twist.)
Simulation of ref. \cite{Cha94} is actually dealing with both magnetic field and
randomness. In the model, however, the randomness is not 
of random chemical potential,
but is of random hopping strength;
the latter may lie
rather out of present scope:
Note that the chemical potential term (particle number)
and the superconductivity order parameter (gauge)
are conjugate variables.
Co-existence of these mutually conjugate terms is
the source of
fascination of this physics. On the contrary, it
causes the aforementioned technical complications.

Here, in order to overcome the difficulty of, so to speak,
`complex-sign problem,'
we employed the exact-diagonalization method.
This method
has not been used so frequently in the course of the studies on this issue.
Perhaps, the limitation of the tractable system sizes had 
been worried over.
Yet, recent extensive simulations and subsequent careful data analyses
are guaranteeing its validity and usefulness \cite{Runge92,Nishiyama99}.
In particular, the exact diagonalization has an advantage over others
that the method gives
dynamical response functions \cite{Gagliano87}
such as the AC conductivity
(without resorting to
the maximum-entropy method, for instance).
Of course, it does manage the complex Hamiltonian elements.

The rest of this paper is organized as follows.
In the next section, we explain our model Hamiltonian
of the two-dimensional hard-core-boson model with magnetic field
and random chemical potential.
Details how we had chosen the gauge will be explicated there.
In Section \ref{section3}, we present numerical results.
For the first time, with use of the finite-size-scaling method,
we observe clear evidences of the onset of
field-tuned localization transition.
After determining the transition point,
we evaluate experimentally accessible quantities such
as the critical conductivity and the dynamical critical exponent.
The last section is devoted to summary and discussions.

\section{Model Hamiltonian --- two-dimensional hard-core-boson model}
\label{section2}

In this section, we explain our model Hamiltonian and
some of its physical ingredients.
The model we simulated is the hard-core-boson model
on the square lattice subjected to external magnetic field
and quenched random chemical potential.
The Hamiltonian is given by,
\begin{equation}
\label{Hamiltonian}
{\cal H}= -\frac{J}{2}
          \sum_{\langle ij \rangle} 
            \left( {\rm e}^{{\rm i}\phi_{ij}} b^\dagger_i b_j 
                 + {\rm e}^{-{\rm i}\phi_{ij}} b^\dagger_j b_i \right)
           + \sum_i H_i  (2b^\dagger_i b_i-1) .
\end{equation}
$\sum_{\langle ij \rangle}$ denotes the summation over all nearest neighbors
on the square lattice of size $L \times L$.
Periodic-boundary condition is imposed as for the horizontal ($x$-axis)
direction.
The operators $\{ b_i \}$ ($\{ b^\dagger_i \}$) are the 
hard-core-boson annihilation (creation) operators at site $i$.
The gauge twist angles $\{ \phi_{ij} \}$ are chosen such as
$\phi_{ij}=(i_y-(L+1)/2) \pi B$ for the horizontal
$i$-$j$ pair of $j$ right next to $i$,
and
$\phi_{ij}=0$ for the vertical $i$-$j$ pair.
Because of this choice of gauge,
a particle hopping clockwise around a plaquette
feels, as a whole,
the gauge angle of $\pi B$.
That is,
the bosons are subjected to the
uniform magnetic flux $\pi B$ per plaquette.
Hereafter, we call $B$ `magnetic field.'
(Therefore, the situation of half flux quantum per plaquette
is realized at the magnetic-field strength $B=1$.)
Quenched random chemical potentials $\{ H_i \}$ distribute
over the sector $[ -\sqrt3 \Delta, \sqrt3 \Delta]$ uniformly
(rectangular distribution).
Therefore, the strength of randomness is 
controlled by the parameter $\Delta=[ H_i^2 ]_{\rm av}$, where
$[ \cdots ]_{\rm av}$ denotes the random-sample average.

Let us try to transcribe the situation realized by the Hamiltonian
(\ref{Hamiltonian}) in a manner appealing our intuition  more vividly:
Hard-core bosons are confined in a
square lattice. 
Throughout this paper, we fix the particle density to be half-filled.
In the absence of $B$ and $\Delta$, 
a statistical mechanical theorem ensures that the
particles are superconducting (namely, the gauge symmetry is broken
spontaneously)
at the ground state \cite{Kishi89};
note that because of the hard-core restriction, 
the onset of superconductivity
is a subtle issue.
The random chemical potential $\{ H_i \}$ may
work so as to disturb the superconductivity.
The strength of the randomness is preferably slightly less than 
that of the localization-transition threshold,
 and so the bosons are left still superconducting.
Because the horizontal periodic boundary condition is imposed,
the lattice is rolled up to form a cylinder.
From each end of the cylinder,
two bar magnets are inserted so that
the cylinder is subjected to magnetic field 
perpendicular to the cylinder surface.
This magnetic field is supposed to destroy the superconductivity 
so that Anderson localization may occur
at a certain critical magnetic field eventually.
This magnetic-field-swept transition 
has not yet been simulated, whereas the randomness-driven transition
has already been simulated extensively 
\cite{Runge92,Makivic93,Batrouni93,Wallin94,Sorensen92,Zhang95,Nishiyama99}.
For a summary of preceding simulation results at $B=0$,
readers may consult with Introduction of ref.
\cite{Nishiyama99}.

It is meaningful to transform the boson Hamiltonian (\ref{Hamiltonian}) 
into the form of
quantum $XX$ model.
With use of the mapping relations between hard-core
boson and $S=1/2$ spin \cite{Matsubara56},
\begin{eqnarray}
b_i^\dagger &=& S^+_i   \nonumber \\
b_i &=& S^-_i   ,
\end{eqnarray}
the above Hamiltonian (\ref{Hamiltonian}) 
is expressed in terms of the $S=1/2$ spin operators,
\begin{equation}
\label{Hamiltonian_spin}
{\cal H}= -\frac{J}{2}
          \sum_{\langle ij \rangle} 
            \left( {\rm e}^{{\rm i}\phi_{ij}} S^+_i S^-_j 
                 + {\rm e}^{-{\rm i}\phi_{ij}} S^-_i S^+_j \right)
           + 2 \sum_i H_i S^z_i  ,
\end{equation}
apart from a constant term.
Now, we are in a position to address some notions
gained from this spin representation:
The onset of superconductivity, that is, the spontaneous breaking
of gauge symmetry, is interpreted as the appearance of 
the in-plane spontaneous magnetization; that is,
\begin{equation}
\label{in-plane_magnetization}
M_{XY}^2 = \sum_{i \ne j} \left(
                   S^x_i S^x_j+S^y_i S^y_j \right)  .
\end{equation}
(In the above, the spin-spin correlations over the identical sites are
subtracted, because they just yield a constant term.
That subtraction improves
finite-size-scaling behavior.)
It is proved rigorously \cite{Kishi89}
that the in-plane symmetry is spontaneous broken
at the ground state of the two-dimensional $XX$ model.
Therefore, it is expected that at least in the vicinity of
$B=0$ and $\Delta=0$, 
our system (\ref{Hamiltonian}) may continue to be
in a superconductor phase.
The perturbations of $B$ and $\Delta$ are supposed to
destroy the superconductivity.
In the spin language, $B$ introduces in-plane magnetic frustration at
each plaquette;
experts at the quantum Monte-Carlo technique may be convinced now why
the method does not work. 
On the other hand,
$\Delta$ appears to be random magnetic field
coupling the $z$-component of spin.
In the next section, we will study how $M_{XY}$ gets disturbed
by these perturbations numerically.

\section{Numerical results}
\label{section3}
In this section, we present numerical results.
We have performed exact-diagonalization calculation
for the Hamiltonian (\ref{Hamiltonian_spin}) with
system sizes up to $L=5$.
We have fixed the particle density to be half-filled
$n(=N/L^2)=0.5$.
For those system sizes of odd
$L$, 
we carried out two sets of simulations for the particle
numbers of $N=[L^2/2]$ and $[L^2/2]-1$
(the bracket $[\cdots]$ denotes the Gauss notation).
Thereby, we performed interpolation calculation (least-square fitting) 
with respect to those data
so as to obtain the data at $n=0.5$.
In the interpolation, we used the relation
$Q(n)=a(n-0.5)^2+c$, which means
that the
physics is symmetric with respect to $n=0.5$ 
(owing to the particle-hole symmetry). 
The amount of randomness $\Delta$ is kept unchanged throughout the simulation;
namely,
$\Delta=0.7$.

\subsection{Field-tuned localization}
\label{subsection3.1}

In this subsection, we will show evidences of the 
superconductor-insulator transition tuned by the magnetic field.
In the analysis, the spin language (\ref{Hamiltonian_spin}) 
is used.
Namely, the localization transition is identified as the disappearance of
the $XY$ (in-plane) magnetic order $M_{XY}$.

\begin{figure}[htbp]
\begin{center}\leavevmode
\epsfxsize=17cm
\epsfbox{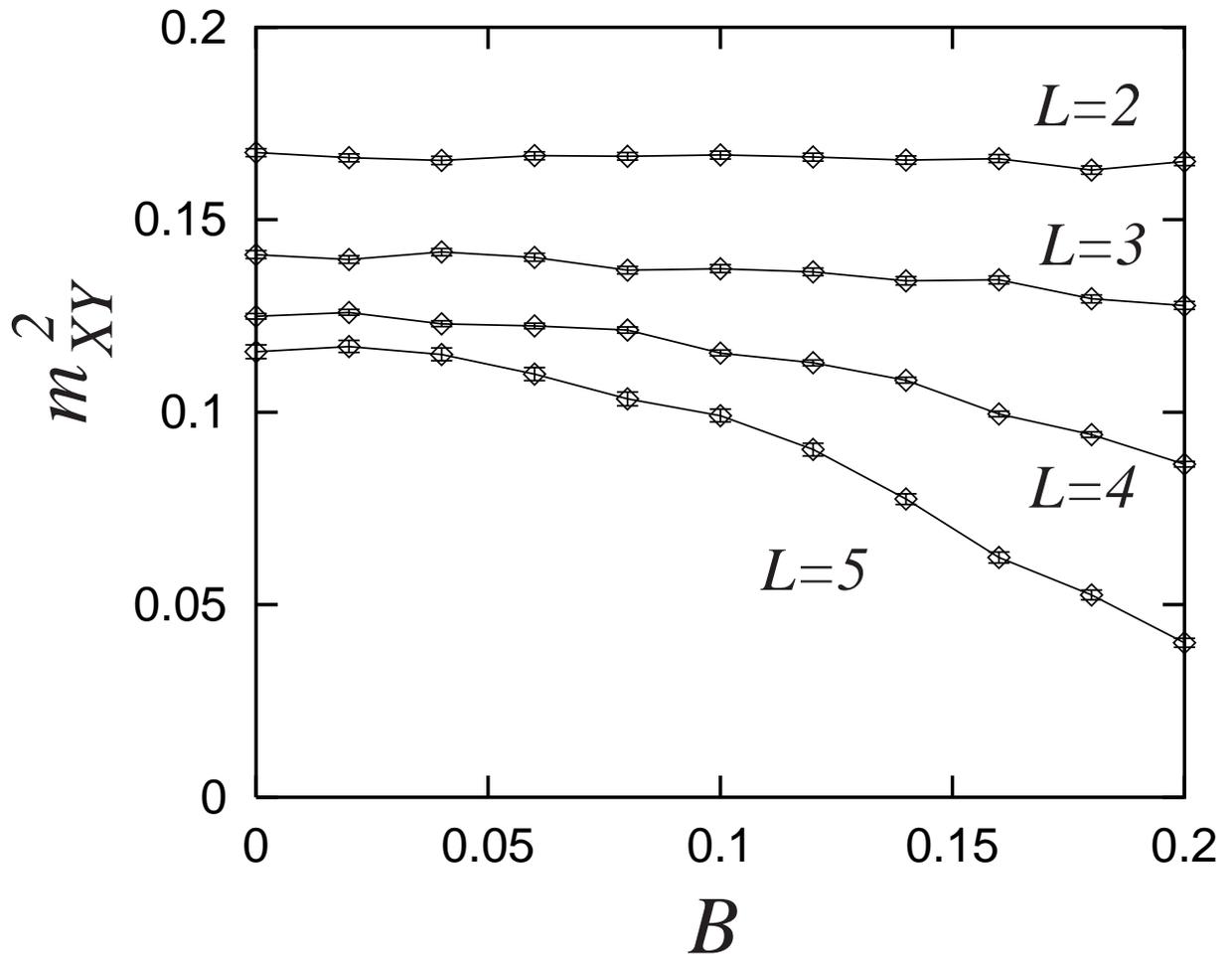}
\end{center}
\caption{
Square of the in-plane magnetization
(superconductivity-order parameter) 
({\protect \ref{in-plane_magnetization}}) is
plotted for $\Delta=0.7$ and various $B$.
From the plot, we see that the in-plane order gets disturbed gradually by 
the magnetic field $B$.
}
\label{Fig_magnetization}
\end{figure}

In Fig. \ref{Fig_magnetization}, we plotted the square of the
in-plane magnetization per site
$m^2=[\langle M_{XY}^2\rangle ]_{\rm av}/L^4$
for $\Delta=0.7$ and various $B$.
The bracket $[\cdots]_{\rm av}$ denotes the random-sample average.
$\langle \cdots \rangle$ denotes the 
ground-state expectation for respective random samples.
The random-sample numbers are $1024$, $1024$, $1024$ and $192$ for 
$L=2$, $3$, $4$ and $5$, respectively.
As is noted above, for odd $L$, we have performed two sets of simulations.
Hence, the number of random samples is effectively twice as many as that
indicated above for odd $L$.
From the plot,
we see that 
the in-plane magnetization gets suppressed by $B$.
However, onset of localization transition seems to be less transparent.
Indication of the localization transition will be seized in the
following analyses.

\begin{figure}[htbp]
\begin{center}\leavevmode
\epsfxsize=17cm
\epsfbox{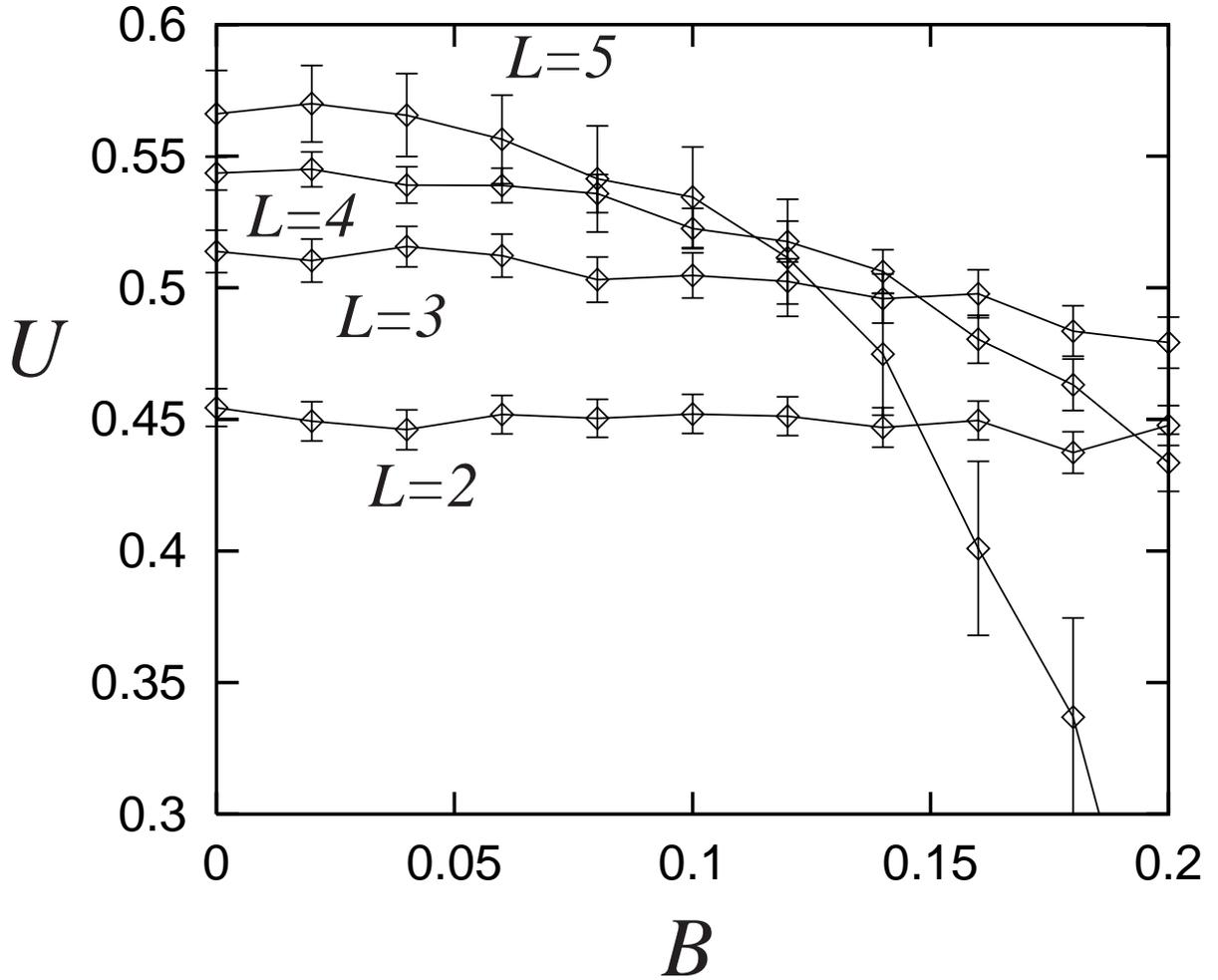}
\end{center}
\caption{
Binder parameter for the in-plane magnetic order 
({\protect \ref{Binder_parameter}})
is plotted for the same parameter range as that of
Fig. {\protect \ref{Fig_magnetization}}.
The intersection point of the curves gives the location of the
localization-transition point.
Hence, we found that at $B \approx 0.1$, localization transition
takes place. 
}
\label{Fig_Binder_parameter}
\end{figure}

In Fig. \ref{Fig_Binder_parameter}, 
we plotted the Binder parameter $U$ \cite{Binder81} for the in-plane
magnetic order; that is,
\begin{equation}
\label{Binder_parameter}
U=1-\frac{ [ \langle M_{XY}^4 \rangle ]_{\rm av} }
             { 3  [ \langle M_{XY}^2 \rangle ]_{\rm av}^2  } .
\end{equation}
Finite-size-scaling behavior of 
the Binder parameter contains information whether $M_{XY}$ develops or not:
As $L$ is enlarged,
the Binder parameter should grow (be suppressed),
provided that
the order
$M_{XY}$ is long-ranged (short ranged).
At critical point,
it remains scale-invariant with respect to $L$.
That is, intersection point of the Binder-parameter curves 
yields the location of the transition point.

From the plot, we see that an intersection point locates 
at $B \approx 0.1$.
Namely, the superconductor phase survives up to the critical magnetic
field $B \approx 0.1$, at which the field-tuned localization 
transition takes place eventually.
However,
rather large corrections to finite-size scaling,
especially for small $L$, prevent precise
determination of the critical point.

\begin{figure}[htbp]
\begin{center}\leavevmode
\epsfxsize=17cm
\epsfbox{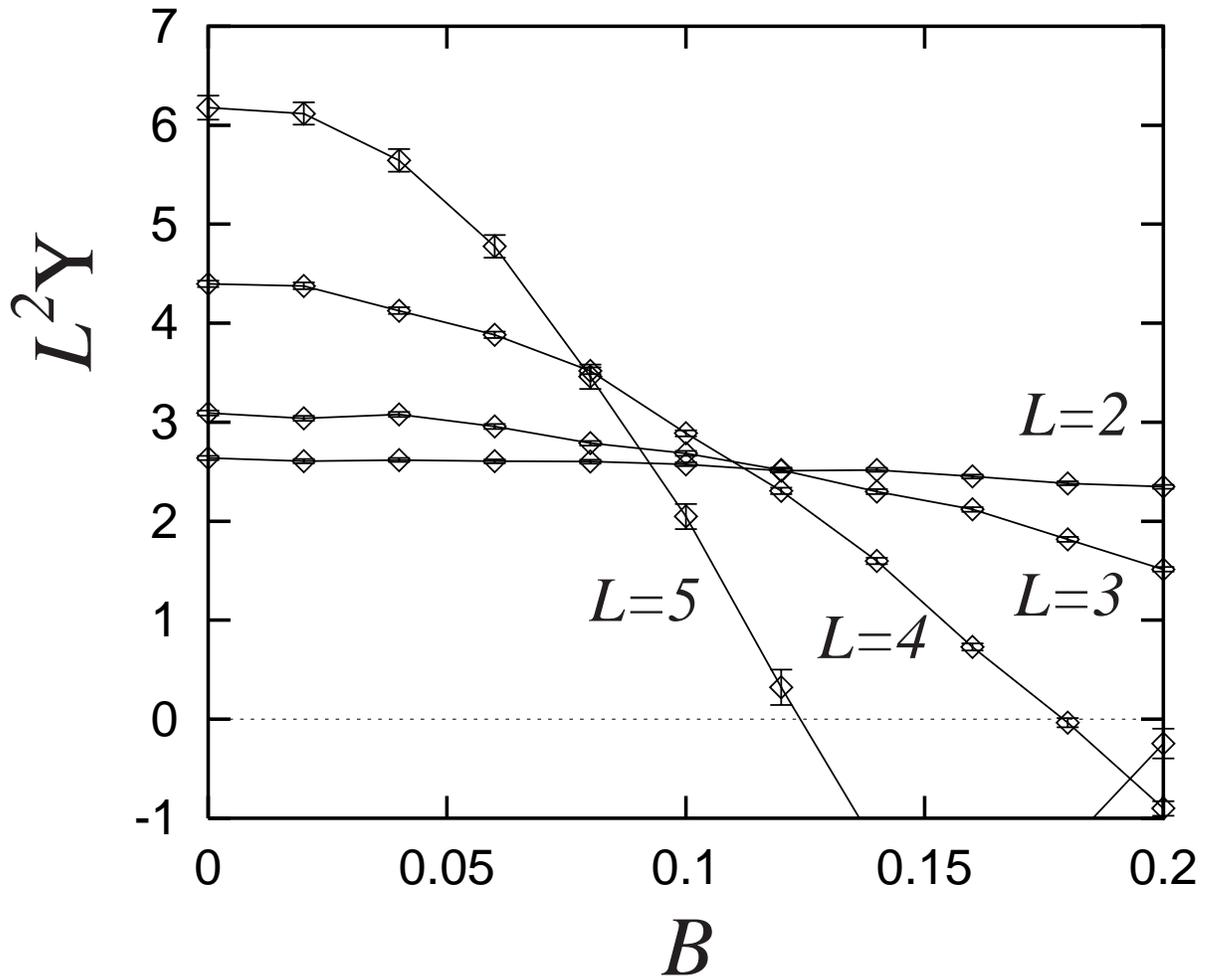}
\end{center}
\caption{
Scaled gauge stiffness ({\protect \ref{stiffness}}) is plotted for the
same parameter range as that of Fig. {\protect \ref{Fig_magnetization}}.
The intersection point of the curves yields the critical point.
Therefore, the estimate $B\approx0.1$ of the former analysis of Fig.
{\protect \ref{Fig_Binder_parameter}} is supported.
In addition, it is suggested that $z=2$ holds.
This relation will be confirmed in the subsequent analysis of Fig.
{\protect \ref{Fig_DelE}}.
}
\label{Fig_scaled_stiffness}
\end{figure}

We carried out an alternative analysis so as to
complement the above observation.
We calculated the gauge stiffness $\Upsilon$ \cite{Fisher73},
which is defined by the
elasticity 
with respect to the boundary gauge twist;
\begin{equation}
\label{stiffness}
\Upsilon = 
\left[
\left\langle
\left.
 \frac{\partial^2 E_{\rm g}(\Theta)}{\partial \Theta^2}
\right|_{\Theta=0}
\right\rangle
\right]_{\rm av} ,
\end{equation}
where $\Theta$ denotes the gauge-twist angle through the
periodic boundary,
and $E_{\rm g} (\Theta)$ is the ground-state
energy under the twisted boundary condition $\Theta$. 
In other words, $\Upsilon$ is the elasticity modulus
against the introduction of
a magnetic flux penetrating through the cylinder;
note that our square lattice is rolled up to form a cylinder.
Therefore,
the stiffness should remain finite in the superconductor phase,
while it vanishes in the Anderson-localization phase.
In Fig. \ref{Fig_scaled_stiffness}, we plotted the scaled stiffness 
$L^2\Upsilon$ for the same parameter range as that of Figs.
\ref{Fig_magnetization} and \ref{Fig_Binder_parameter}.
We observe that the stiffness is suppressed by $B$ as would be expected.
In order to gain further information from this scaling plot,
we have to consult with the
formal scaling argument on $\Upsilon$ \cite{Fisher89,Fisher90a}.
According to this argument,
at the localization point, the stiffness should obey the scaling relation
$\Upsilon \sim L^{-(d+z-2)}$.
That is, the scaled stiffness $L^{d+z-2}\Upsilon$ should be scale-invariant
at the localization transition point.
Actually, from the plot, we see that the scaled stiffness with the 
particular choice of $z=2$
exhibits scale invariance at critical point $B\approx0.1$;
the validity of $z=2$ will be confirmed in the analysis
of Section \ref{subsection3.3}.
Putting together the finite-size-scaling behaviors of $U$ and $\Upsilon$,
we are led to the conclusion that the field-tuned localization takes
place at $B\approx0.1$.
In the following, we calculate experimentally accessible quantities
at the point $B=0.1$.

\subsection{Critical conductivity}

Here, we evaluate the AC conductivity at the localization point
$B=0.1$ determined in the above subsection.
As is mentioned in Introduction, an argument claims
that the DC conductivity would be a universal value of 
the order 
$\approx (e^*)^2/h$
with charge of one particle $e^*$ (say, Cooper pair).
Various experiments have tried to validate this issue.
However, the results are
still remaining controversial; see Table \ref{table1}.
Nevertheless, it should be noted that finite conductivity itself
is a novel feature lying out of the scope of the conventional localization theory
in two dimensions; see Introduction.

According to the Kubo formula, the AC conductivity 
is expressed in terms of 
 the current-current correlation function,
\begin{equation}
\sigma(\omega)={\rm Re}
\left[
\frac{1}{\hbar\omega L^2}
 \int_0^\infty {\rm d}t 
  {\rm e}^{{\rm i}\omega t}
   \left\langle
    \left[
     J_x(t),J_x
    \right]
   \right\rangle
\right]_{\rm av} ,
\end{equation}
with the current operator 
$J_x = \frac{{\rm i}e^*J}{2\hbar}
  \sum_{j,\delta_x}  
      {\rm e}^{ {\rm i} \phi_{j,j+\delta_x} } \delta_x
            a_{j+\delta_x}^\dagger a_j$.
This formula reduces to the resolvent form,
\begin{equation}
\sigma(\omega)=
 {\rm Re}
  \left(
  \frac{{\rm i}}{\hbar\omega L^2}
   \left[
    \left\langle
     J_x
      \left(
       \frac{\hbar}{E_{\rm g}-{\cal H}+\hbar\omega +{\rm i}\eta}
      +\frac{\hbar}{E_{\rm g}-{\cal H}-\hbar\omega -{\rm i}\eta}
      \right)
     J_x
    \right\rangle
   \right]_{\rm av}
  \right)  .
\end{equation}
Hence, one is forced to calculate the inverse of the Hamiltonian matrix,
which is seemingly impossible in practice.
However, Gagliano and Balseiro found that the resolvent is expanded
into the following
 continued-fraction form \cite{Gagliano87},
\begin{equation}
\left\langle
f_0
\left|
 \frac{1}{z-{\cal H}}
\right|
f_0
\right\rangle
 =
\frac{\langle f_0|f_0\rangle}{
 z-\alpha_0-\frac{\beta_1^2}{
 z-\alpha_1-\frac{\beta_2^2}{\ddots}
                            }
                            }        ,
\end{equation}
with the coefficients generated by the following recursion relations,
\begin{eqnarray}
 |f_{i+1}\rangle &=& {\cal H}|f_i\rangle -\alpha_i|f_i\rangle
                                      -\beta_i^2|f_{i-1}\rangle , \nonumber \\
        \alpha_i &=& \langle f_i |{\cal H}| f_i \rangle 
                               / \langle f_i | f_i \rangle  ,  \nonumber \\
      \beta_i^2 &=& \langle f_i | f_i \rangle 
                              /\langle f_{i-1} |f_{i-1}\rangle
                                       \ \ (\beta_0=0) .
\end{eqnarray}
These procedures are essentially the same as those of the
Lanczos tri-diagonalization.
It is one of major advantages of the Lanczos diagonalization
that one can calculate dynamical response functions.

\begin{figure}[htbp]
\begin{center}\leavevmode
\epsfxsize=17cm
\epsfbox{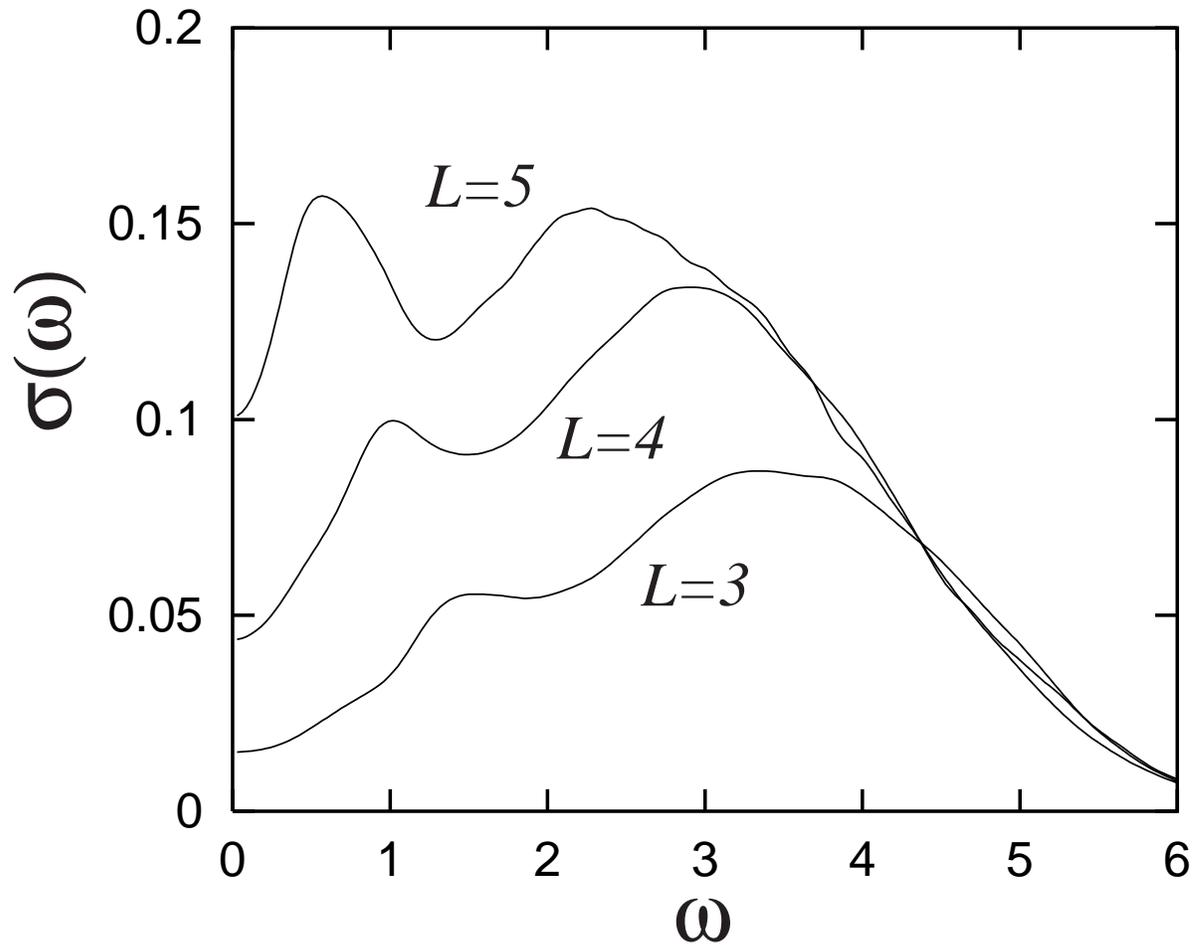}
\end{center}
\caption{
AC conductivity at the localization-transition point ($\Delta=0.7$ and
$B=0.1$) is plotted. Rapid drop in the vicinity of $\omega=0$ is
due to the finite-size effect. 
Hence, the maximal conductivity
beside $\omega=0$ has to be regarded as the DC conductivity for
each $L$.
}
\label{Fig_AC_conductivity}
\end{figure}

\begin{figure}[htbp]
\begin{center}\leavevmode
\epsfxsize=17cm
\epsfbox{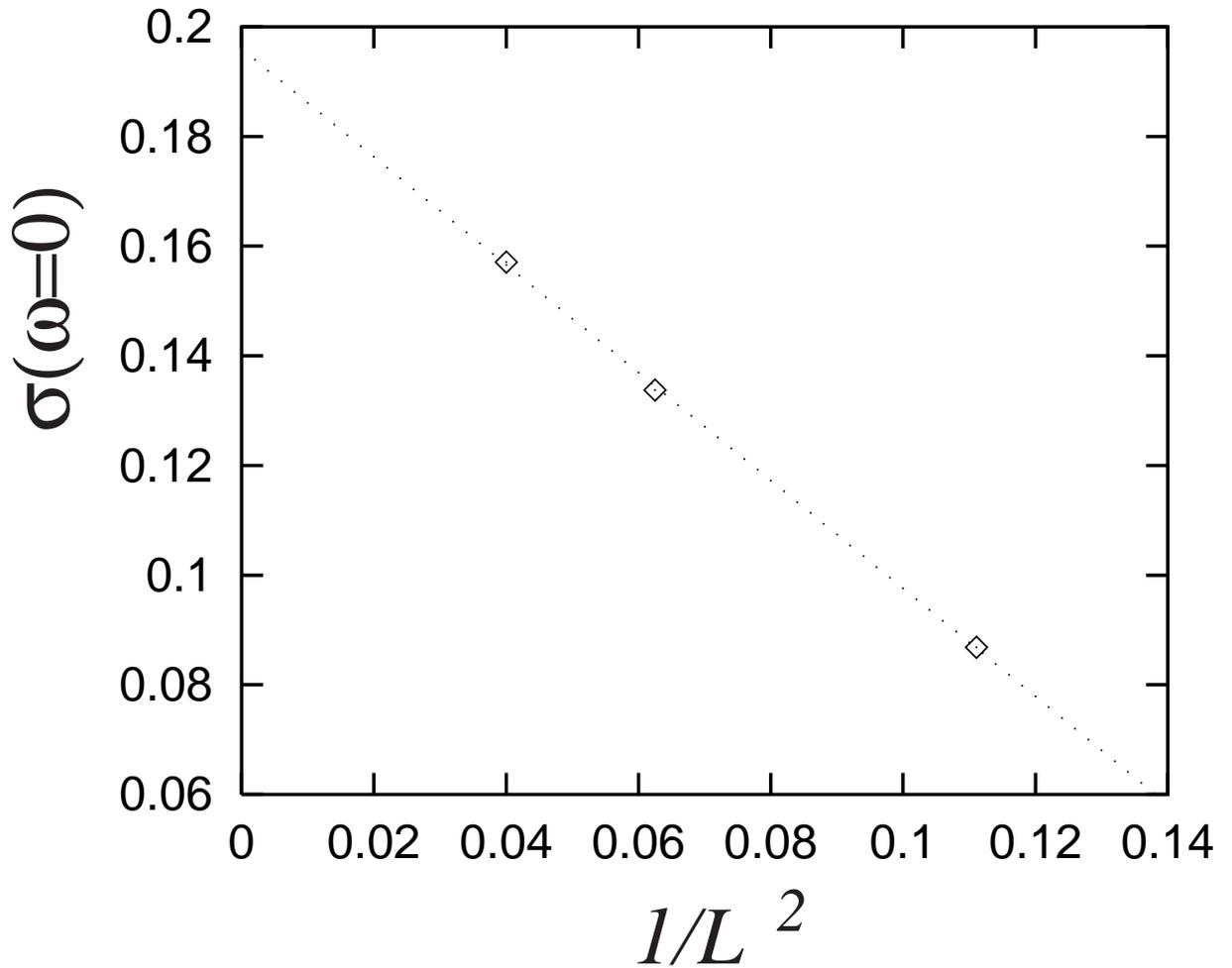}
\end{center}
\caption{
DC conductivity at the localization-transition point ($\Delta=0.7$ and
$B=0.1$) is plotted against $1/L^2$.
Through the least-square fit, we obtained the critical DC conductivity
in the $L\to\infty$ limit as 
$\sigma_{\rm c}=0.196\pm0.01 ((e^*)^2/h)$.
This value turns out to be
substantially larger than that
of the randomness-driven transition $\sigma_{\rm c}\approx0.135 ((e^*)^2/h)$
at $B=0$. 
}
\label{Fig_DC_conductivity}
\end{figure}

In Fig. \ref{Fig_AC_conductivity}, 
we plotted the AC conductivity at the transition point ($B=0.1$ and $\Delta=0.7$);
we had set $\eta=0.2$.
The data are averaged over random samples of $1024$, $1024$ and
$64$ for $L=3$, $4$ and $5$, respectively.
Because of finite-size energy gap above the ground state,
the AC conductivity in Fig. \ref{Fig_AC_conductivity} 
drops in the vicinity of $\omega=0$.
AC conductivity curve forms a peak beside $\omega=0$, and
as $L$ is enlarged,
the peak position shifts toward the center
$\omega=0$.
Hence,
for sufficiently large system sizes,
the drop of $\sigma(\omega\approx0)$ may disappear, and
a Drude-like peak centered at $\omega=0$ may emerge instead.
Therefore, we regard the maximal conductivity beside $\omega=0$
as the
DC conductivity for respective $L$.
Since the DC conductivity for each $L$ exhibits finite-size correction,
we need to extrapolate those finite-$L$ data
to the value of
thermodynamic limit.
In Fig. \ref{Fig_DC_conductivity}, we plotted the finite-$L$ DC conductivity 
against $1/L^2$; the choice of this abscissa scale is due to
the reasoning addressed in the literature \cite{Runge92}.
Through the least-square fit, we obtained the DC conductivity
in the thermodynamic limit
such as $\sigma_{\rm c}=0.196\pm0.001 ((e^*)^2/h)$. 
Remarkably enough,
we notice that this conductivity is larger than 
that of the randomness-driven transition 
$\sigma_{\rm c} \approx 0.135 ((e^*)^2/h)$ at $B=0$ \cite{Nishiyama99},
suggesting breakdown of the universality of $\sigma_{\rm c}$.
We will discuss this point in the last section.

\subsection{Dynamical critical exponent}
\label{subsection3.3}

\begin{figure}[htbp]
\begin{center}\leavevmode
\epsfxsize=17cm
\epsfbox{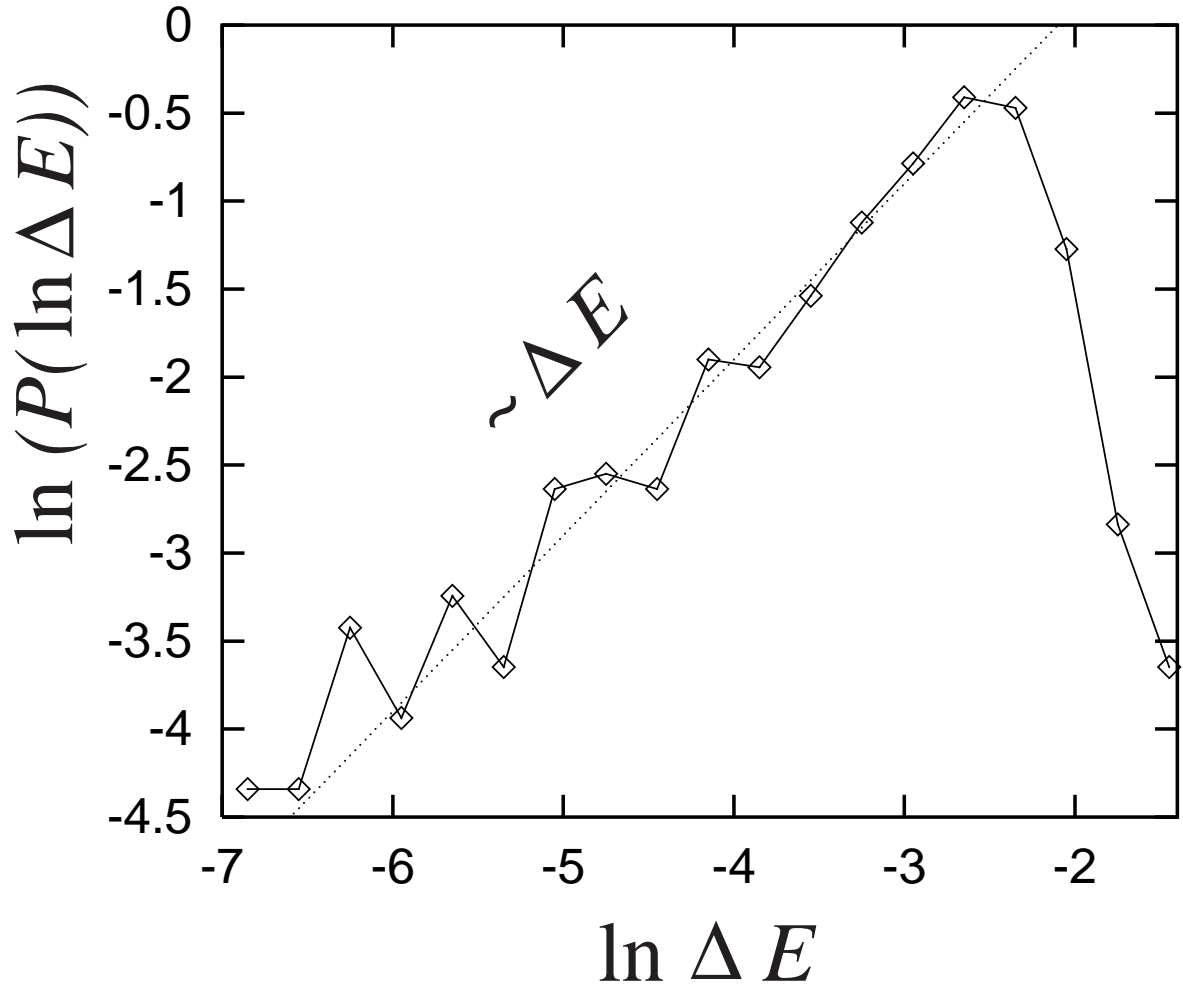}
\end{center}
\caption{Probability distribution of $\log$ of the first energy gap 
over $512$ random samples with $L=5$
at the localization-transition point ($\Delta=0.7$ and $B=0.1$).
The dotted line shows the slope of $P \sim \Delta E$.
From the low-energy tail of this distribution, 
we obtain an estimate of
the dynamical critical exponent $z=2$.
}
\label{Fig_DelE}
\end{figure}

In the subsection \ref{subsection3.1},
we have already used the relation
$z=2$ for the purpose of investigating scaling property of
the gauge stiffness $\Upsilon$
(\ref{stiffness}).
In this subsection, we obtain more conclusive estimate
of $z$ with use of the
Rieger-Young relation \cite{Rieger96}.
In Fig. \ref{Fig_DelE}, we show the probability distribution
of the logarithm of the first energy gap $\log \Delta E$ over
$512$ random samples with $L=5$
at the transition point ($B=0.1$ and $\Delta=0.7$).
According to the Rieger-Young argument,
the low-energy tail of the distribution contains an information
of the dynamical critical exponent.
In the following, we recollect their argument briefly.
Because our system is disordered, the first energy gap $\Delta E$
distributes obeying a certain probability-distribution function $P$.
We are considering critical phenomenon, where the physics is scale-invariant
and no such particular energy scale that characterize the physics
exists.
Therefore, it is sensible to consider $P(\log \Delta E)$ 
rather than $P(\Delta E)$. 
Low-lying excitation may be created at a peculiar part 
of random sample,
where local excitation costs very little energy.
Therefore,
the probability may be proportional to the spatial
volume; namely, $P \propto L^d\Delta E^\lambda$ with the exponent
$\lambda$ describing the low-energy tail.
On the other hand, the finite-size-scaling theory insists that
it should be of the form $P=\tilde{P}(L/\xi_{\rm r})=\tilde{P}(L/\xi_{\tau}^{1/z})$
(definition of $z$).
With use of the relation $\Delta E\sim 1/\xi_{\tau}$,
the distribution turns out to be
a function of $L(\Delta E)^{1/z}$.
Therefore, we obtain the relation,
\begin{equation}
\log P(\log \Delta E) = \lambda \log \Delta E + C,
\end{equation}
with $\lambda=d/z$.

In Fig. \ref{Fig_DelE}, we observe that the low-energy tail is, in fact, governed
by that scaling relation with the exponent
$\lambda=1$.
In consequence, we obtain the relation $z=d=2$, confirming
the preliminary estimate obtained in the subsection \ref{subsection3.1}.

\section{Summary and discussions}
\label{section4}

We have simulated the two-dimensional hard-core-boson model 
(\ref{Hamiltonian})
subjected to the external magnetic field and 
the random chemical potential $\Delta=0.7$.
Difficulty arising from the complex hopping amplitudes
$\{ {\rm e}^{{\rm i}\phi_{ij}} \}$, that have been preventing
the application of the quantum Monte-Carlo method,
is circumvented by the use of
the exact-diagonalization scheme.
For the first time, in computer simulation,
we observed evidences of the 
field-tuned localization transition with
the finite-size-scaling technique applied to
the gauge order $M_{XY}$ as well as the gauge stiffness $\Upsilon$.
After determining the location of the localization point,
we carried out calculations of two experimentally
accessible quantities, namely, the critical conductivity and
the dynamical critical exponent.
Thereby,
we obtained those estimates such as
$\sigma_{\rm c}=0.196\pm0.01((e^*)^2/h)$ and
$z = 2$.
First, let us make a comparison with preceding simulation results.
Although no former results for $B\ne0$ are available, 
there have been reported a number of simulation studies at $B=0$ 
\cite{Runge92,Makivic93,Batrouni93,Wallin94,Sorensen92,Zhang95,Nishiyama99}.
These simulation results are settling gradually
to the conclusions such as
$\sigma_{\rm c} \approx  0.135 ((e^*)^2/h)$ and $z=2$.
Hence, contrary to common belief,
we are led to conclude that the critical conductivity 
is not universal but 
increases as the external magnetic field is strengthened.
As for $z$, on the other hand,
we found that the dynamical critical exponent $z=2$ is
retained even for $B \ne 0$.

Secondly, let us turn our attention to making a comparison with experiments.
As is noted above, the quantities evaluated here 
are measurable in experiments;
a summary of experimental results is presented
in Table \ref{table1}.
Among various reports listed there,
in particular, we would like to draw reader's attention to the
latest exhaustive measurement by Markovi\'c {\it et al.} 
\cite{Markovic98,Markovic99},
who 
scanned both magnetic-field strength and film thickness (randomness).
Their experiment demonstrates that
the critical conductivity (resistivity) does not stay
universal but increases (decreases) gradually as $B$ is applied.
Our numerical result supports this behavior 
\cite{Yazdani95,Markovic98,Markovic99}.
Non-universality of $\sigma_{\rm c}$ has been arousing
much attention. Meanwhile,
there have been proposed several ideas
intending to account for that
\cite{Shimshoni98,Mason99,Wagenblast97,Kramer98,Rimberg97,%
Trivedi90,Huscroft98};
for instance,
it was proposed that one should take into account
dissipative modes (possibly un-paired normal electrons) that
would give rise to extra resistivity in addition to the intrinsic one.
Our result shows, however, that without resorting to such dissipative extra modes,
one can account for the gradual increase of
$\sigma_{\rm c}$ within the framework of the
pure-boson model (\ref{Hamiltonian}). 
As for $z$, it appears that the
numerical simulation result $z=2$ is not in agreement with experiments.
The discrepancy 
had already come out in the past
studies of $B=0$.
There had been proposed an attempt at 
altering the universality class \cite{Wallin94,Sorensen92};
It was claimed that an inclusion of long-range repulsion 
among particles would 
alter the universality class.
Because in our exact-diagonalization calculation,
tractable system sizes might not be sufficient to implement
long-range interaction, it would be exceedingly cumbersome
 to validate this
issue definitely.


In preliminary stage of our simulation, we had swept various parameter 
ranges so as to find optimal condition to observe localization transition:
Note that $\Delta$ should not exceed a threshold at which 
particles localize already at zero magnetic field $B=0$.
On the other hand, $\Delta$ should not be too weak, because
superconductor-insulator transition may occur at
strong magnetic field ($B\sim0.5$). 
For such $B$ in close proximity to some primal fractional numbers,
certain specific type of vortex-lattice structure becomes
favored so that simulation data with sweeping magnetic field
suffer from insystematic behaviors.
Moreover, for large $B$,
one cannot judge at all whether superconductivity is destroyed solely by the magnetic
field or assistance of randomness is also significant.
After sweeping various parameter ranges,
we had 
determined the optimal randomness $\Delta=0.7$, and
thereby, we
performed extensive simulations at this particular condition.
It would be remained for future study to gain inclusive features
for the whole region including strong magnetic field.

\section*{Acknowledgment}
Numerical calculation was performed on Alpha workstations
of theoretical physics group, Okayama university, and
on the vector supercomputers Hitachi SR8000/60, 
of the institute of solid state physics,
university of Tokyo.



\begin{table}
\begin{tabular}{ l  c  c c}
Reference & Sample & Critical exponents & Critical resistivity \\
\hline
Hebard {\it et al.} \cite{Hebard90} & ${\rm InO}_x$ film
                & $z\approx1$, $z\nu \approx 1.3$  & $5000\Omega$  \\
Tanda {\it et al.} \cite{Tanda92}   & High-$T_{\rm c}$ film
                & $z\nu=1.2(1)$                    & $8.5{\rm k}\Omega$ \\
Seidler {\it et al.} \cite{Seidler92} & High-$T_{\rm c}$ film
                & $z\nu=1.37(10)$                  & $13.6{\rm k}\Omega$ \\
Chen {\it et al.} \cite{Chen95}       & JJ array 
                & $z\nu=1.5\sim8$,                 & non-universal \\
                                      &     
                & $z=1.05$                         &               \\
Yazdani {\it et al.} \cite{Yazdani95} & MoGe film 
                & $\nu\approx1.5$, $z=1.0(1)$      
                                   & $R_\Box\downarrow$ as $B_{\rm c}\uparrow$ \\
van der Zant {\it et al.} \cite{Zant96} & JJ array
                & $z\nu=1.2\sim2$,  
                                   & non-universal \\
                                        &
                & $z=0.34\sim1.4$  &                                \\
Markovi\'c {\it et al.} \cite{Markovic98,Markovic99} & Bismuth film
                & $z\approx1$, $\nu\approx0.7$     
                                   & $R_\Box\downarrow$ as $B_{\rm c}\uparrow$ \\
\end{tabular}
\caption{
A summary of previous experimental measurements.
Because even in a single literature, a number of different data are presented
for respective samples, we have presented above a mean value.
}
\label{table1}
\end{table}

\end{document}